\documentclass{emulateapj}
\usepackage{apjfonts}
\usepackage{epsf}

\slugcomment{Accepted in ApJ}

\newcommand\bmu{\mbox{\boldmath $\mu$}}
\newcommand\reffig[1]{Figure~\ref{fig:#1}}

\newcommand\reftab[1]{Table~\ref{tab:#1}}

\begin{document}

\title{Differential Microlensing of the Continuum and Broad Emission Lines
in SDSS J0924+0219, the Most Anomalous Lensed Quasar\footnote{Based on
observations made with the NASA/ESA Hubble Space Telescope, obtained at
the Space Telescope Science Institute, which is operated by AURA, Inc.,
under NASA contract NAS 5-26555.}}

\author{
  Charles R.\ Keeton\altaffilmark{1},
  Scott Burles\altaffilmark{2},
  Paul L.\ Schechter\altaffilmark{2},
  and Joachim Wambsganss\altaffilmark{3}
}

\altaffiltext{1}{
  Department of Physics \& Astronomy, Rutgers University,
  Piscataway, NJ 08837 USA
}
\altaffiltext{2}{
  Kavli Institute for Astrophysics and Space Research and Department
  of Physics, Massachusetts Institute of Technology, Cambridge, MA 02139 USA
}
\altaffiltext{3}{
  Astronomisches Rechen-Institut, Zentrum f\"ur Astronomie,
  Universit\"at Heidelberg, M\"onchhofstrasse 12-14, 69120 Heidelberg,
  Germany
}

\begin{abstract}
SDSS J0924+0219 is the most glaring example of a gravitational lens
with anomalous flux ratios: optical broad-band photometry shows image
D to be a factor of 12 fainter than expected for smooth lens potentials.
We report spectroscopy showing that the anomaly is present in the broad
emission line flux ratios as well.  There are differences between the
emission line and continuum flux ratios: the A/D ratio is 10 in the
broad Lyman-$\alpha$ line and 19 in the associated continuum.  Known
variability argues for the presence of microlensing.  We show that
microlensing can account for both the continuum and emission line
flux ratios, if the broad emission line region is comparable in size
to the Einstein radii of the microlenses.  Specifically, we need the
half-light radius of the broad-line region to be
$R_{\rm BLR} \lesssim 0.4\,R_E \sim 9$ lt-days, which is small but
reasonable.  If the broad-line region is that large, then stars can
contribute only 15--20\% of the surface mass density at the positions
of the images.  While we cannot exclude the possibility that
millilensing by dark matter substructure is present as well, we
conclude that microlensing is present and sufficient to explain
existing data.  Under this hypothesis, the A/D flux ratio should
return to a value close to unity on a time scale of years rather than
millennia.
\end{abstract}


\section{Introduction}

The gravitational lens system SDSS J0924+0219 presents a fascinating
challenge for lens modelers.  Discovered by \citet{inada0924}, images
from the Sloan Digital Sky Survey showed what appeared to be a triple
system in a very odd configuration.  The lensing galaxy could be seen
in higher resolution follow-up observations obtained with the Baade
telescope at Las Campanas Observatory, which in combination with the
three known images called for a fourth image.  Modeling and subtracting
the three bright images (A, B, and C) did reveal a fourth image, D, but
it was a factor of 10 fainter than predicted.  The image positions are
typical of an ``inclined quad'' or ``fold'' configuration produced by
a source near a fold caustic \citep[see][]{saha}, so images A and D
ought to be nearly equal in brightness \citep[see][]{foldreln}.

Anomalies in lens flux ratios are possible when there is small-scale
structure in the lensing galaxy, in the form of either dark matter
subhalos \citep[millilensing; e.g.,][]{MS,MM,chiba,DK} or stars
\citep[microlensing; e.g.,][]{CR,SW}.  The challenge is to understand
whether either possibility can actually explain why image D is so
faint.  One might hope to discriminate between milli- and microlensing
by observing a component of the QSO that is large compared to the
Einstein rings of stars but small compared to the Einstein rings of
subhalos \citep{moustakas,wisotzki0435,metcalf2237,chibaIR,wayth}.
We therefore obtained broad emission line flux ratios for
SDSS J0924+0219 using the Hubble Space Telescope.  We describe our
observations in \S 2, our analysis in \S 3, and our conclusions in
\S 4.  We assume a cosmology with $\Omega_M = 0.3$,
$\Omega_\Lambda = 0.7$, and $H_0 = 70$ km s$^{-1}$ Mpc$^{-1}$.
The redshift of the source quasar is $z_s = 1.524$ \citep{inada0924},
and the redshift of the lens galaxy is $z_l = 0.394$ \citep{eigenbrod}.

\section{Observations}

\subsection{Imaging}

\begin{figure*}[t]
\plottwo{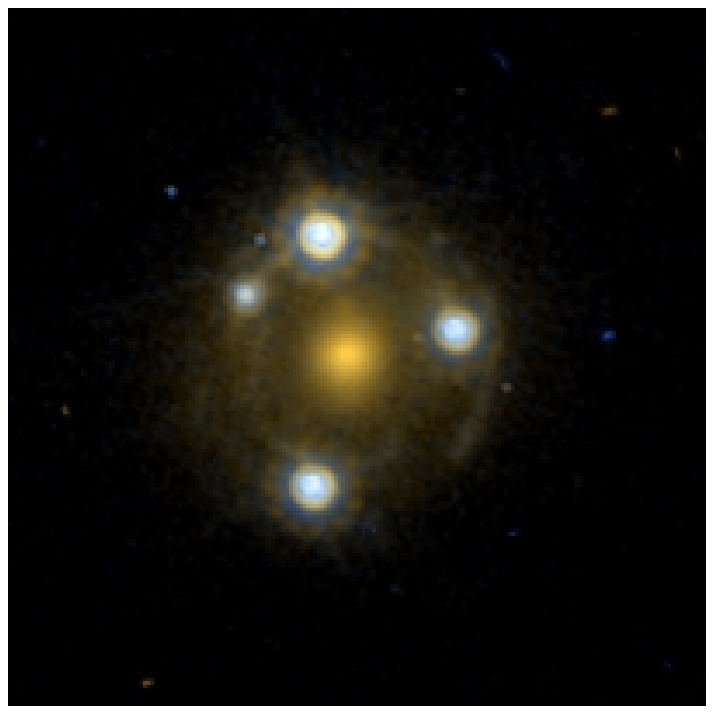}{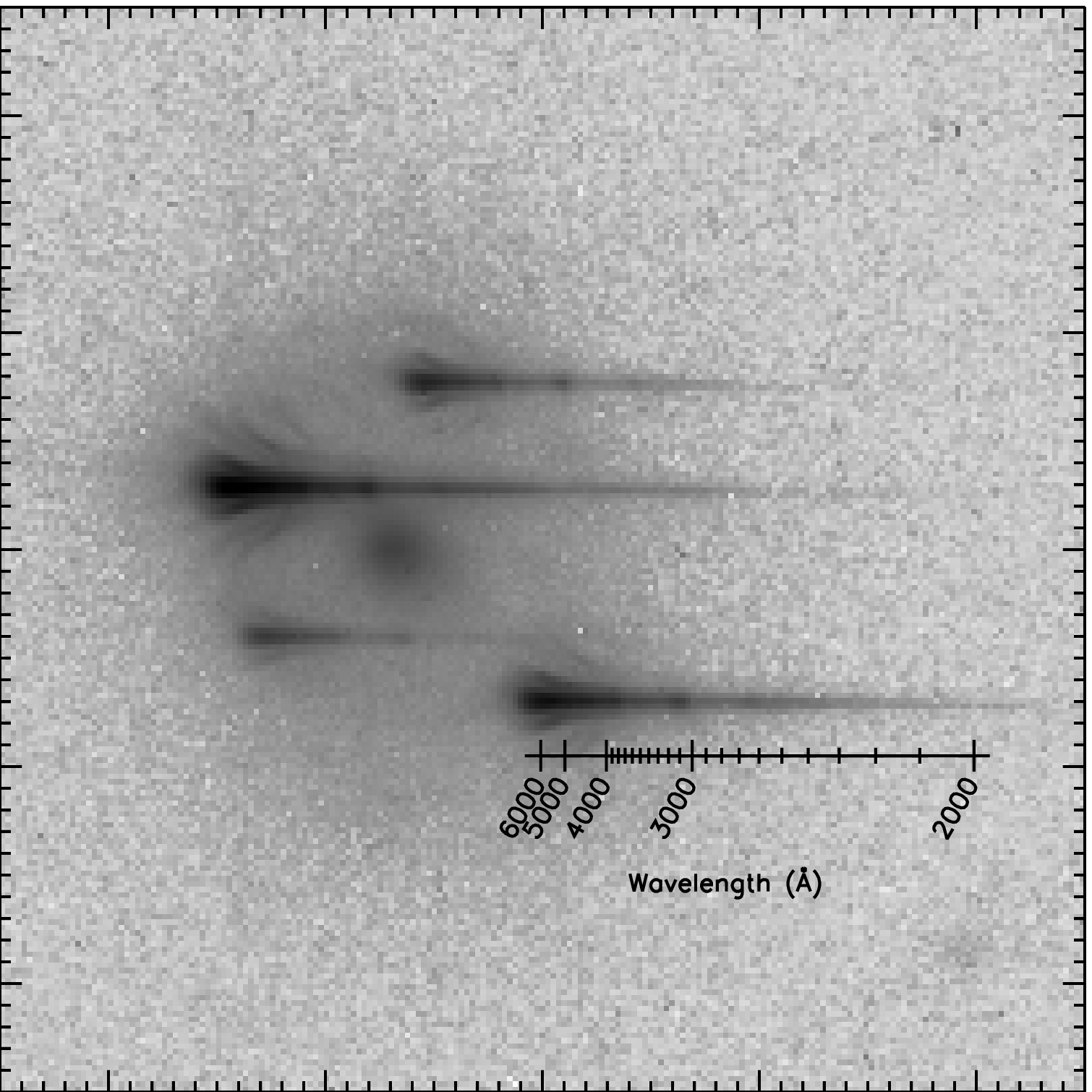}
\caption{
Direct ACS/WFC (left) and dispersed ACS/HRC (right)
5\arcsec$\times$5\arcsec\ images of SDSS J0924+0219 from the Hubble
Space Telescope.  The direct image is oriented with North up and East
to the left.  Clockwise from the top, the quasar components are
labeled A, C, B, and D.  The dispersed image is rotated by
$\sim$70\arcdeg.  Each image is displayed with a non-linear stretch
given by an inverse hyperbolic sine to highlight both bright and
faint features.  An approximate wavelength scale is superposed on
the dispersed image to show the effective wavelength solution.
}\label{fig:acs}
\end{figure*}

SDSS0924 was observed with the Wide Field Channel (WFC) of the
Advanced Camera for Surveys (ACS) on the Hubble Space Telescope, as
part of program GO-9744 (PI\ C.\ Kochanek).  Four 547-sec exposures
were obtained with the F555W filter ($\approx$V) on 18 November 2003,
and four 574-sec exposures were obtained with the F814W ($\approx$I)
on 19 November 2003.  An interpolated color composite of the V- and
I-band images is shown in \reffig{acs}a.  The four QSO components and
the lens galaxy are clearly visible.  An Einstein ring image of the
QSO host galaxy is also apparent in the I-band image.

We performed a non-linear least squares fit to the flattened frames
from standard ACS data processing, using a photometric model
consisting of four point sources and a de Vaucouleurs model galaxy,
convolved with a TinyTim PSF \citep[][v6.2]{krist}.  We fit each
exposure separately and used the scatter among exposures to assess
the measurement uncertainties.  For the galaxy we find an I-band
effective radius $R_{\rm eff} = 0\farcs436 \pm 0\farcs004$,
an axis ratio $q = 0.92\pm0.02$, and a position angle
$\theta_q = -25\arcdeg \pm 5\arcdeg$ (East of North).  The
positions and broad-band fluxes of the QSO components are listed
in \reftab{data}.  The HST flux ratios are consistent with those
reported by \citet{inada0924}.

\begin{deluxetable*}{crrcrrrrr}
\tablewidth{0pt}
\tablecaption{Component positions and fluxes of SDSS J0924+0219}
\tablehead{
  \colhead{Image} &
  \colhead{$-\Delta\alpha\ (\arcsec)$} &
  \colhead{$ \Delta\delta\ (\arcsec)$} &
  \colhead{$\sigma\ (\arcsec)$} &
  \colhead{$F_{I}$} &
  \colhead{$F_{V}$} &
  \colhead{$F_{17}$} &
  \colhead{$F_{12}$} &
  \colhead{$F_{\alpha}$}
}
\startdata
 A & $ 0.0000$ & $ 0.0000$ & 0.0010 & $529.5\pm3.2$ & $274.9\pm2.5$ & 5.59 & 2.49 & 120.3 \\
 B & $-0.0636$ & $-1.8063$ & 0.0010 & $246.4\pm2.1$ & $141.8\pm0.7$ & 1.66 & 0.66 & 46.2 \\
 C & $ 0.9648$ & $-0.6788$ & 0.0013 & $181.3\pm0.7$ & $106.7\pm1.5$ & 0.62 & 0.24 & 21.6 \\
 D & $-0.5414$ & $-0.4296$ & 0.0026 & $ 36.7\pm0.5$ & $ 16.9\pm0.6$ & 0.29 & 0.10 & 11.8 \\
 G & $ 0.1804$ & $-0.8685$ & 0.0039 & $376.8\pm1.5$ & $ 61.3\pm2.9$ &      &       &
\enddata
\tablecomments{
$F_I$ and $F_V$ are broad-band fluxes in the HST I and V bands,
respectively, expressed in counts per second.  The uncertainties
derived from the scatter among exposures are consistent with
Poisson noise.  These fluxes can be converted to AB magnitudes
using zeropoints of 25.937 in I and 25.718 in V.  $F_{17}$ and
$F_{12}$ are spectroscopic continuum fluxes at rest wavelengths
of 1700 \AA\ and 1216 \AA, respectively, in units of
$10^{-17}$ ergs cm$^{-2}$ s$^{-1}$ \AA$^{-1}$.  Component D has
a systematic uncertainty of $\sim$20\% in $F_{12}$ due to galaxy
subtraction (see text).  $F_\alpha$ is the flux in the
Ly$\alpha$/\ion{N}{5} broad emission line, in units of
$10^{-17}$ ergs cm$^{-2}$ s$^{-1}$.
}
\label{tab:data}
\end{deluxetable*}

\subsection{Spectroscopy}

We planned for SDSS0924 to be observed as the final and most
important target in our HST program GO-9854 to obtain spatially
resolved spectroscopy of eight quadruply lensed systems with the
Space Telescope Imaging Spectrograph (STIS).  The observations were
attempted on 19 June 2004, but guide star acquisition failed.  We
arranged to reobserve SDSS0924 with the same setup, but the power
supply to STIS was shut down before the new observations were
carried out.

We considered whether ACS could be used to conduct the desired 
observations, and realized that the sapphire prism (PR200L) in the
High Resolution Channel (HRC) of ACS would provide coverage of the
broad emission lines of Lyman-$\alpha$, \ion{Si}{4} and \ion{C}{4}.
The observations were successfully executed on 29 May 2005.  The
combined image of three 880-sec exposures is shown in \reffig{acs}b.
The visible and near-UV light from 6000 \AA\ to 2000 \AA\ is dispersed
over 80 native HRC pixels.  The dispersion per pixel ranges from
14000 km s$^{-1}$ at 6000 \AA\ to 4000 km s$^{-1}$ at 3000 \AA\ to
1300 km s$^{-1}$ at 2000 \AA.

\begin{figure}[t]
%%\epsscale{0.7}
\plotone{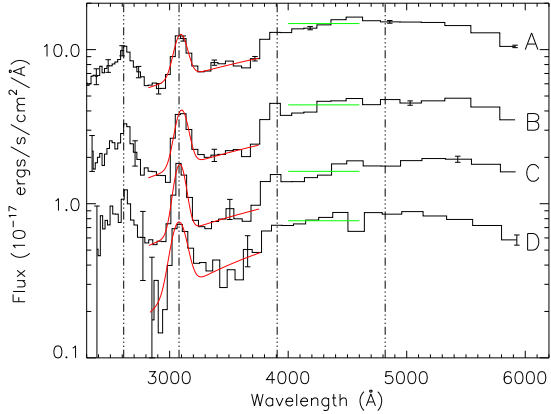}
\caption{
Extracted spectra of the four images, versus observed wavelength.
Sample errorbars are shown only in certain bins for clarity.  The
vertical lines indicate the expected locations of the broad emission
line complexes Ly$\beta$/\ion{O}{6}, Ly$\alpha$/\ion{N}{5},
\ion{C}{4}, and \ion{C}{3}], given the quasar redshift $z_s = 1.524$
\citep{inada0924}; they provide an external check on our wavelength
solutions.  The red curves show our models of Gaussian emission
lines at Ly$\alpha$/\ion{N}{5}, with linear continua, fit to each
spectrum over the rest wavelength range 1120--1460 \AA.  The green
lines indicate the average continuum levels over the rest wavelength
range 1600--1800 \AA.
}\label{fig:spec}
\end{figure}

The combined cosmic ray rejected product from STScI provided the
starting point for spectral extraction.  We subtracted the galaxy by
masking the dispersed footprints of the four QSO images and modeling
the galaxy surface brightness distribution with a linear b-spline
model \citep[cf.][]{bolton}.  The relative wavelength was given by
the ground calibration of ACS \citep{pavlovsky}, and we shifted the
wavelength scale of each spectrum to match the obvious emission
features (Ly$\alpha$/\ion{N}{5}, Ly$\beta$/\ion{O}{6}, \ion{C}{4})
present in the QSO dispersed images.  The spectrum of each QSO image
was constructed by a simple boxcar extraction in each column with a
boxcar half-width in pixels given by $\sqrt{\lambda}/24$.  The count
rate in each spectral pixel was converted to flux using sensitivity
tables produced from ground calibrations.  The extracted spectra are
shown in \reffig{spec}.

We fit a simple five parameter model composed of a linear continuum
and a Gaussian emission line to each spectrum over the range
1120--1460 \AA\ in the QSO frame, covering the Ly$\alpha$/\ion{N}{5}
broad emission line.  The fits are shown in \reffig{spec}, and the
line and continuum fluxes are listed in \reftab{data} as $F_\alpha$
and $F_{12}$, respectively.   We also report the average continuum
flux level over the rest wavelength range 1600--1800 \AA\ as $F_{17}$.

\reffig{acs}b shows that the spectra for images A and D pass fairly
close to the center of the galaxy.  To assess systematic effects due
to the galaxy subtraction, we repeated the measurements with both 5\%
more and less galaxy subtracted.  All of the spectral flux measurements
in \reftab{data} varied by less than 5\%, with the exception of the
continuum measurement $F_{12}$ for image D which varied by 20\%.  One
other possible systematic effect is that $F_{17}$ may contain a small
amount ($\lesssim$5\%) of broad line flux, but that should not affect
our conclusions.

\subsection{Comments}

SDSS0924 is anomalous not just in broad-band photometry but also in the
continuum and broad emission line flux ratios.  Moreover, the anomalies
are different in different passbands: $A/D$ is 10 in the emission line,
14--16 in the broad-band filters, and $>$19 in the continuum.  That
the broad-band values lie between the emission line and continuum
values makes sense if the broad-band filters contain both continuum
and emission line light.  (We estimate that $\sim$10\% of the V-band
flux is from broad emission lines, while the I-band contains $\sim$20\%
emission line flux including the 3000 \AA\ bump.)  For comparison,
the values for $A/C$ are 5.6 in the emission line, 2.6--2.9 in the
broad-band filters, and $>$9 in the continuum.  That the broad-band
values lie below both the emission line and continuum values may seem
puzzling.  However, we must recall that the photometry and spectroscopy
come from different epochs, and image C has faded since November 2003
\citep{csk0924} such that the broad-band ratio is presently $A/C \sim 5$.

\section{Analysis}

\subsection{Basic picture}

Optical broad-band photometry shows that the images vary independently
and on a time scale of years \citep{csk0924}, which is long compared
with the predicted time delays (see \S 3.2) and therefore implies that
microlensing is present.  Microlensing can also explain why the emission
line and continuum flux ratios differ, as we shall demonstrate
\citep[also see][]{schneider,metcalf2237,wayth}.

What remains is to explain why the emission line flux ratios are
anomalous.  We initially suspected millilensing, but then realized
that we would need at least two clumps since both C and D have
anomalous emission line flux ratios.  Since we know that microlensing
is present, a simpler hypothesis is that microlensing produces all
of the anomalies.  We will show that microlensing can indeed explain
the data for SDSS0924.

One obvious question is whether microlensing has the right scale to
affect broad emission lines.  The source plane Einstein radius of a
star of mass $M$ is
\begin{equation}
  R_E = \left( \frac{4 G M}{c^2}\ \frac{D_{ls} D_{os}}{D_{ol}} \right)^{1/2}
  \approx 22\,\left(\frac{M}{M_\odot}\right)^{1/2}\mbox{ lt-days}.
\end{equation}
For comparison, reverberation mapping studies have shown that the
broad line regions of active galactic nuclei span a range of sizes
but certainly extend down to $\sim$10 lt-days and below
\citep[e.g.,][and references therein]{kaspi}.  The comparison is
not perfect because reverberation mapping studies have focused on
Balmer lines (especially H$\beta$) while we have observed Ly$\alpha$,
and because reverberation mapping and microlensing are probably
sensitive to source geometry in different ways.  Nevertheless, it
appears that the microlensing scale is not grossly inappropriate.

\subsection{Macromodel}

For a microlensing analysis we need to estimate the convergence and
shear at the position of each image.  We treat the lens galaxy as a
singular isothermal ellipsoid whose surface mass density (in units
of the critical density for lensing) can be written as
\begin{equation}
  \frac{\Sigma(r,\theta)}{\Sigma_{\rm crit}} = \frac{b}{2r}
    \left[\frac{1+q^2}{(1+q^2) - (1-q^2)\cos 2(\theta-\theta_q)}\right]^{1/2} ,
\end{equation}
where $q$ is the projected axis ratio, $\theta_q$ is the orientation
angle, and $b$ is a mass parameter related to the Einstein radius of
the lens.  We constrain the center of the mass distribution using the
observed galaxy position.  We optimize $b$ and $q$, but fix $\theta_q$
to match the orientation of the observed galaxy \citep[see][]{CSKlight}.
We include an external tidal shear to allow for the possibility that
the environment of the lens galaxy (which is currently unknown) affects
the lens potential \citep[e.g.,][]{KKS}.

We fit the image and galaxy positions (not the flux ratios, since they
are the subject of the microlensing analysis).  The best-fit model has
$\chi^2 = 33$ for $\nu = 2$ degrees of freedom, which is dominated by
the galaxy position suggesting that we may have underestimated its
uncertainty.  (Eliminating the galaxy position constraint leads to
$\chi^2 = 2$.)  Allowing the orientation to be free would yield
$\chi^2 = 15$ for $\nu = 1$, again dominated by the galaxy position,
but this model has a misalignment of $48\arcdeg$ between the mass and
the light so we deem it to be implausible.

The model has an axis ratio $q_{\rm mod} = 0.81$, which is not very
different from the observed axis ratio $q_{\rm obs} = 0.92$.  It
has an external shear $\gamma_{\rm ext} = 0.10$ at position angle
$\theta_{\rm ext} = 71\arcdeg$ (East of North); such a shear is
typical for four-image lenses and suggests that the lens may lie in
a modest group of galaxies \citep[e.g.,][]{momcheva}.  The predicted
convergence $\kappa$, shear $\gamma$, and magnification $\mu$ for
each image are listed in \reftab{kapgam}.  The predicted time delays
between the images are all less than 12 days.  Thus, independent
variability in the images on a time scale of years cannot be
attributed to intrinsic variability of the source.

\begin{deluxetable}{crrr}
\tablewidth{0pt}
\tablecaption{Macromodel properties}
\tablehead{
  \colhead{Image} &
  \colhead{$\kappa$} &
  \colhead{$\gamma$} &
  \colhead{$\mu$}
}
\startdata
 A & 0.502 & 0.458 & $ 26.2$ \\
 B & 0.503 & 0.405 & $ 12.0$ \\
 C & 0.511 & 0.560 & $-13.4$ \\
 D & 0.476 & 0.565 & $-22.4$
\enddata
\label{tab:kapgam}
\end{deluxetable}

\subsection{Toy microlensing}

A toy model for microlensing features a single star that is nearly
aligned with one of the lensed images.  The image will be split into
several ``microimages'' separated by a distance comparable to the star's
Einstein radius $R_E$.  On this tiny scale the effects of the galaxy
can be approximated as a constant convergence $\kappa$ and shear
$\gamma$, so that the lens equation can be written as
\begin{eqnarray}
  u &=& x \left( 1 - \kappa - \gamma - \frac{R_E^2}{x^2+y^2} \right) , \\
  v &=& y \left( 1 - \kappa + \gamma - \frac{R_E^2}{x^2+y^2} \right) ,
\end{eqnarray}
in coordinates centered on the star and aligned with the shear.
For a point source, the magnification of a microimage at $(x,y)$ is
\begin{equation}
  \mu(x,y) = \left\{ (1-\kappa)^2 - \gamma^2
    - \frac{R_E^2 [R_E^2 - 2\gamma(x^2-y^2)]}{(x^2+y^2)^2}
    \right\}^{-1} .
\end{equation}

Suppose the star is perfectly aligned with the macroimage, so
$u = v = 0$.  Consider a negative-parity macroimage, such that
$1-\kappa-\gamma<0$ and $1-\kappa+\gamma>0$.  Then the lens equation
is easily solved to find that there are two microimages at $x = 0$
and $y = \pm R_E/\sqrt{1-\kappa+\gamma}$, and each has magnification
$\mu = -1/[4\gamma(1-\kappa+\gamma)]$.\footnote{If the macroimage has
positive parity (such that $1-\kappa+\gamma > 1-\kappa-\gamma > 0$),
there are two additional microimages at
$x = \pm R_E/\sqrt{1-\kappa-\gamma}$ and $y = 0$, each having
magnification $\mu = 1/[4\gamma(1-\kappa-\gamma)]$.}  Since the
microimages are not separately resolved, what matters is the
combined magnification
$\mu_{\rm tot} = 1/[2\gamma(1-\kappa+\gamma)]$ \citep[also see][]{SW}.
Using the convergence and shear values from \reftab{kapgam} yields
$\mu_{\rm tot} = -0.813$, which corresponds to a flux ratio
$D/A = 0.031$.

For imperfect alignment, we can still solve the lens equation
analytically if the source lies on the $u$-axis or $v$-axis (see
Appendix B.1 of \citealt{cuspreln} for details).  For a source
lying a distance $d < R_E$ from the origin along either axis, we
find $D/A = 0.031 + 0.027(d/R_E)^2 + {\mathcal O}(d/R_E)^4$.

Clearly microlensing can cause image D to be very faint, if there is
a star reasonably close to the macroimage.

\subsection{Realistic microlensing}

Going beyond the toy model, we face three important questions:
(1) When the density of stars is low (so their caustics do not
intersect), how likely is it that there is a star close enough to
the macroimage to produce strong flux perturbations?
(2) As the density of stars increases (so their caustics merge),
is it still possible to get a strong demagnification?
(3) How do the conclusions differ for point-like and finite sources?

\begin{figure}[t]
%%\epsscale{0.7}
\plotone{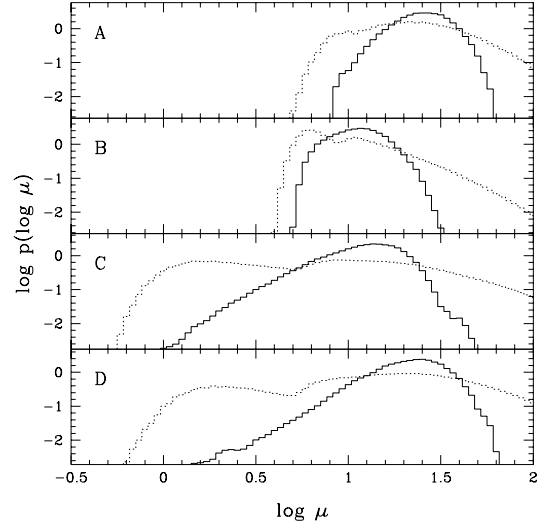}
\caption{
Microlensing magnification distributions for the four images of
SDSS0924, assuming $f_* = 15\%$ of the surface mass density is in
stars.  The dotted histograms show the distributions for the continuum,
assuming it to be a point source.  The solid histograms show the
results for the emission lines, assuming the broad line region to be
a Gaussian with a half-light radius $R_{\rm BLR}/R_E = 0.4$.
}\label{fig:magdist}
\end{figure}

To answer these questions, we run microlensing simulations using the
ray shooting software by \citet{w90,wps,jcam}.  We use the convergences
and shears from \reftab{kapgam}, and assume that a fraction $f_*$ of
the surface mass density is in stars.  The software produces a
magnification map for each image.  Treating the continuum region as
a point source, we can make a histogram of the magnifications in the
map to obtain the probability distribution for the continuum
magnification.  Given a model of the broad line region, we can
convolve the magnification map with that model before making the
histogram to obtain the probability distribution for the emission
line magnification.  We treat the BLR as a Gaussian with half-light
radius $R_{\rm BLR}$, since \citet{mortonson} argue that details other
than the half-light radius do not significantly affect microlensing
magnification distributions.  Examples of the distributions are shown
in \reffig{magdist}.

To analyze both continuum and emission line flux ratios, we must
determine the joint probability distribution $p(\mu_c,\mu_l)$ for
the continuum and emission line magnifications of each image.  For
each source position, we take $\mu_c$ from the raw magnification
map and $\mu_l$ from the convolved map, and then use all pixels to
construct a histogram in the $(\mu_c, \mu_l)$ plane.

Finally, to simultaneously analyze images A and D we compute the
four-dimensional probability distribution\footnote{To simplify the
notation, we collect the four magnifications into the vector
$\bmu = (\mu^A_c, \mu^A_l, \mu^D_c, \mu^D_l)$.}
\begin{equation}
  p(\bmu) = p_A(\mu^A_c,\mu^A_l) \times p_D(\mu^D_c,\mu^D_l)\,.
\end{equation}
This represents the probability density for image A to have continuum
and emission line magnifications of $\mu^A_c$ and $\mu^A_l$,
respectively, while image D has $\mu^D_c$ and $\mu^D_l$.  The joint
four-dimensional distribution is just the product of the two
two-dimensional distributions because microlensing is independent
in the two images.

To quantify microlensing's ability to explain our data for SDSS0924,
we then compute
\begin{equation}
  P = \int d\bmu\ p(\bmu) \times \cases{
      1 & $\mu^A_c/\mu^D_c \ge 19$ and $\mu^A_l/\mu^D_l \ge 10$ \cr
      0 & else
    }
\end{equation}
This represents the probability that, if we picked a four-image
lens at random, it would be at least as anomalous as what we have
observed.

This is not really the right figure of merit for evaluating our
data, though, because we did not select SDSS0924 at random; we chose
it specifically because it is the most anomalous of the 22 known
four-image lenses.  A better figure of merit is the probability
$P_N$ of finding at least one strong anomaly in a sample of $N$
four-image lenses.  Unfortunately, it is not clear how to compute
$P_N$.  For $N$ identical lenses, the binomial distribution would
give $P_N = 1 - (1-P)^N \approx N P$ (for $P \ll 1$).  The known
lenses are not identical, but short of doing microlensing
simulations for all of them (interesting, certainly, but beyond
the scope of this work) we must assume that the value of $P$ for
SDSS0924 is representative.  Next, we must determine an appropriate
value for $N$.  There are 16 four-image lenses with ``fold'' or
``cusp'' configurations, in which a close pair or triplet of images
can be analyzed in a robust, model-independent way to find flux
ratio anomalies \citep{cuspreln,foldreln}.  We therefore set $N=16$,
and consider this to be somewhat conservative since we are
neglecting the non-fold/cusp images in these 16 lenses as well as
all images in the six known ``cross'' lenses.  Lastly, we must decide
what value of $P_N$ represents a reasonable threshold of acceptability.
Since we are dealing with complicated posterior probabilities we
should avoid being too demanding, so we believe it is reasonable to
consider $P_N = 1\%$, or $P \approx 0.01/16 \approx 6.25 \times 10^{-4}$.

\begin{figure}[t]
%%\epsscale{0.7}
\plotone{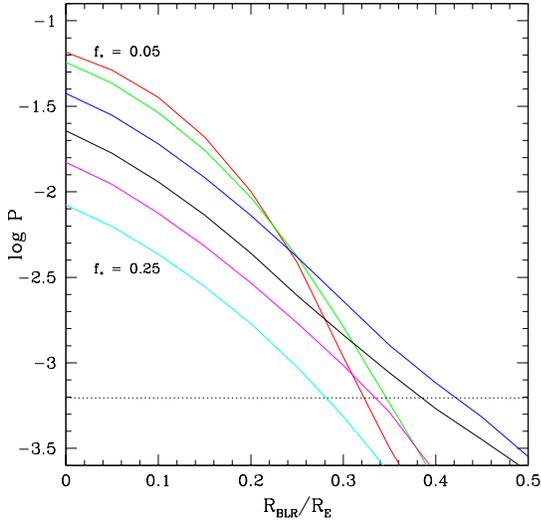}
\caption{
Logarithm of the $P$-value, or the probability of having a flux ratio
anomaly more extreme than observed, as a function of the half-light
radius of the broad line region.  Each curve has a fixed value of
the stellar mass fraction $f_*$; at left, the curves are in order
$f_* = 0.05, 0.10, 0.15, 0.20, 0.25$ from top to bottom.  The dotted
line shows our nominal acceptability threshold $P = 6.25 \times 10^{-4}$.
}\label{fig:prob}
\end{figure}

\reffig{prob} shows $P$ as a function of the half-light radius of
the broad-line region, for different values of the stellar mass
fraction $f_*$.  There are acceptable models with the BLR as large
as $R_{\rm BLR}/R_E \sim 0.4$, for a stellar mass fraction of
$f_* \sim 15$--20\%.  A larger broad-line region is ruled out by
these models.  A smaller broad-line region would allow a wider range
of stellar mass fractions.  Thus, we conclude that there is a range
of microlensing models with reasonable parameters that can explain
the flux ratio anomaly in SDSS0924 acceptably well.

There are two limitations to our present analysis.  First, we assumed
for simplicity that the stellar mass fraction is the same for images
A and D.  Second, we assumed that all the stars have the same mass.
The stellar mass function has very little effect on the magnification
distribution for a point source \citep[see][and references therein]{SWL}.
The same result cannot hold for an extended source, because a large
source must not be sensitive to stars that are sufficiently small,
but the problem of microlensing of a finite source by unequal-mass
stars has not yet been studied.  Relaxing our two assumptions should
only make it {\em easier} to find microlensing models that fit the
data, which means that they do not affect our main conclusion that
microlensing can provide a reasonable explanation for SDSS0924.

\section{Discussion}

We have discovered that the continuum and broad emission line flux
ratios in SDSS0924 differ from each other at the factor of two level
(for both images C and D).  That fact, together with photometric
variability \citep{csk0924}, establishes that microlensing is present
in this system.  We have also found that image D is highly anomalous
in both the continuum and the broad emission line fluxes.  We have
shown that all of these results can be explained by microlensing.

The key point is that saddle images can be strongly suppressed by
microlensing.  The suppression is generally greater for a point
source than for an extended source, which is why images C and D are
fainter (relative to A) in the continuum than in the emission line.
Even so, microlensing can produce a factor of 10 suppression in the
emission line flux of image D, provided that the QSO broad emission
line region has a half-light radius
$R_{\rm BLR} \lesssim 0.4\,R_E \sim 9$ lt-days.  While this result
suggests that $R_{\rm BLR}$ is smaller than we thought broad-line
regions to be when we began our project \citep[also see][]{moustakas},
it seems reasonable in light of results from reverberation mapping of
active galactic nuclei \citep[e.g.,][]{kaspi}.  That is all we can
say without a detailed understanding of geometric factors that may
make sizes measured from microlensing different from those measured
by reverberation mapping.  But it is enough for our proof of principle
that microlensing is sufficient to explain all current data for
SDSS0924.

We believe that microlensing offers the best and most natural
explanation for SDSS0924, but concede that we cannot rigorously rule
millilensing out.  Testing whether millilensing is present will require
new data, such as flux ratios in narrow emission lines (the narrow
line region is generally thought to be large; e.g., \citealt{kraemer};
but see \citealt{bennert} for a contrasting view) or mid-infrared
photometry \citep[e.g.,][]{chibaIR}.  Another intriguing possibility
is spectroscopic variability.  If there is only microlensing, then
over the next few years image D ought to return to the brightness
predicted by smooth lens models --- in both the continuum and emission
line.  If there is any millilensing, then differences between observed
and smooth model flux ratios will persist for centuries.

\acknowledgements
We are deeply grateful to Galina Soutchkova and David Solderblom at
STScI for rescheduling our observations of SDSS0924 and supporting our
request for ACS prism observations.
We thank Chris Kochanek and the other members of HST program GO-9744
for arranging the direct ACS images.
We thank A.\ Eigenbrod for communicating his measurement of the lens
galaxy redshift prior to publication.
We thank Frederic Courbin for valuable comments on the manuscript.
We thank the referee, Prasenjit Saha, for constructive criticism and
for helpful suggestions about the probability analysis.
Support for HST program GO-9854 was provided by NASA through a grant
from STScI, which is operated by the AURA, Inc., under NASA contract
NAS 5-26555.

%\clearpage


\begin{thebibliography}{}

\bibitem[Bennert et al.(2002)]{bennert}
Bennert, N., Falcke, H., Schulz, H., Wilson, A. S., \& Wills, B. J.
2002, \apj, 574, L105

\bibitem[Bolton et al.(2005)]{bolton}
Bolton, A. S., Burles, S., Koopmans, L. V. E., Treu, T., \&
Moustakas, L. A. 2005, in preparation

\bibitem[Chang \& Refsdal(1979)]{CR}
Chang, K., \& Refsdal, S. 1979, \nat, 282, 561

\bibitem[Chiba(2002)]{chiba}
Chiba, M. 2002, \apj, 565, 17

\bibitem[Chiba et al.(2005)]{chibaIR}
Chiba, M., Minezaki, T., Kashikawa, N., Kataza, H., \& Inoue, K. T.
2005, astro-ph/0503487

\bibitem[Dalal \& Kochanek(2002)]{DK}
Dalal, N., \& Kochanek, C. S. 2002, \apj, 572, 25

\bibitem[Eigenbrod et al.(2005)]{eigenbrod}
Eigenbrod, A., et al. 2005, \aap, submitted

\bibitem[Inada et al.(2003)]{inada0924}
Inada, N., et al. 2003, \aj, 126, 666

\bibitem[Kaspi et al.(2005)]{kaspi}
Kaspi, S., Moaz, D., Netzer, H., Peterson, B. M., Vestergaard, M., \&
Jannuzi, B. T. 2005, astro-ph/0504484

\bibitem[Keeton et al.(2003)Keeton, Gaudi \& Petters]{cuspreln}
Keeton, C. R., Gaudi, B. S., \& Petters, A. O. 2003, \apj, 598, 138

\bibitem[Keeton et al.(2005)Keeton, Gaudi, \& Petters]{foldreln}
Keeton, C. R., Gaudi, B. S., \& Petters, A. O. 2005, astro-ph/0503452

\bibitem[Keeton et al.(1997)Keeton, Kochanek \& Seljak]{KKS}
Keeton, C. R., Kochanek, C. S., \& Seljak, U. 1997, \apj, 482, 604

\bibitem[Kochanek(2002)]{CSKlight}
Kochanek, C. S. 2002, in The Shapes of Galaxies and Their Dark Matter
Halos, ed. P. Natarajan (Singapore: World Scientific), 62

\bibitem[Kochanek(2004)]{csk0924}
Kochanek, C. S. 2004, in The Impact of Gravitational Lensing on
Cosmology (IAU 225), eds. Y. Mellier \& G. Meylan; also astro-ph/0412089

\bibitem[Kraemer et al.(1998)]{kraemer}
Kraemer, S. B., Crenshaw, D. M., Filippenko, A. V., \& Peterson, B. M.
1998, \apj, 499, 719

\bibitem[Krist(1995)]{krist}
Krist, J. 1995, ASP Conference Series, 77, 349

\bibitem[Mao \& Schneider(1998)]{MS}
Mao, S., \& Schneider, P. 1998, \mnras, 295, 587

\bibitem[Metcalf \& Madau(2001)]{MM}
Metcalf, R. B., \& Madau, P.. 2001, \apj, 563, 9

\bibitem[Metcalf et al.(2004)]{metcalf2237}
Metcalf, R. B., Moustakas, L. A., Bunker, A. J., \& Parry, I. R.
2004, \apj, 607, 43

\bibitem[Momcheva et al.(2005)]{momcheva}
Momcheva, I., Williams, K., Keeton, C., \& Zabludoff, A. 2005, \apj, submitted

\bibitem[Mortonson et al.(2005)]{mortonson}
Mortonson, M. J., Schechter, P. L., \& Wambsganss, J. 2005, \apj, 628, 594

\bibitem[Moustakas \& Metcalf(2003)]{moustakas}
Moustakas, L. A., \& Metcalf, R. B. 2003, \mnras, 339, 607

\bibitem[Pavlovsky et al.(2004)]{pavlovsky}
Pavlovsky, C., et al. 2004, "ACS Instrument Handbook", Version 5.0,
(Baltimore: STScI).

\bibitem[Saha \& Williams(2003)]{saha}
Saha, P., \& Williams, L. L. R. 2003, \aj, 125, 2769 

\bibitem[Schechter \& Wambsganss(2002)]{SW}
Schechter, P. L., \& Wambsganss, J. 2002, \apj, 580, 685

\bibitem[Schechter et al.(2004)]{SWL}
Schechter, P. L., Wambsganss, J., \& Lewis, G. F. 2004, \apj, 613, 77

\bibitem[Schneider \& Wambsganss(1990)]{schneider}
Schneider, P., \& Wambsganss, J. 1990, \aap,, 237, 42

\bibitem[Wambsganss(1990a)]{w90}
Wambsganss, J. 1990a, PhD Thesis (Munich), also MPA report 550

\bibitem[Wambsganss(1990b)]{wps}
Wambsganss, J., Paczy\'nski, B., \& Schneider, P. 1990b, ApJ, 358, L33

\bibitem[Wambsganss(1999)]{jcam}
Wambsganss, J. 1999, JCAM, 109, 353

\bibitem[Wayth et al.(2005)]{wayth}
Wayth, R. B., O'Dowd, M., \& Webster, R. L. 2005, \mnras, 359, 561

\bibitem[Wisotzki et al.(2003)]{wisotzki0435}
Wisotzki, L., Becker, T., Christensen, L., Helms, A. Jahnke, K., Kelz, A.,
Roth, M. M., \& Sanchez, S. F. 2003, \aap, 408, 455

\end{thebibliography}
\end{document}